\documentclass[%
 aip,
 amsmath,amssymb,
 groupedaddress, 
 reprint,%
]{revtex4-1}

\usepackage{graphicx}% Include figure files
\usepackage{dcolumn}% Align table columns on decimal point
\usepackage{bm}% bold math
\usepackage{hyperref}

\usepackage{bm}
\newcommand{\bb}{\bm{b}}
\newcommand{\brho}{\bm{\rho}}

\newcommand{\bomega}{\bm{\omega}}

\newcommand{\bD}{\bm{D}}
\newcommand{\be}{\bm{e}}

\newcommand{\bk}{\bm{k}}
\newcommand{\bx}{\bm{x}}
\newcommand{\bX}{\bm{X}}

\newcommand{\bU}{\bm{U}}
\newcommand{\bF}{\bm{F}}
\newcommand{\bH}{\bm{H}}
\newcommand{\bq}{\bm{q}}
\newcommand{\bQ}{\bm{Q}}

\newcommand{\bI}{\bm{I}}
\newcommand{\bL}{\bm{L}}

\newcommand{\odiff}{\text{d}} 
\newcommand{\avg}[1]{\left< #1 \right>}
\newcommand{\bavg}[1]{\big\langle #1 \big\rangle}
\newcommand{\Bavg}[1]{\Big\langle #1 \Big\rangle}
\newcommand{\bn}{\bm{n}}
\newcommand{\bj}{\bm{j}}
\usepackage{xcolor}

\usepackage[utf8]{inputenc}
\usepackage[T1]{fontenc}
\usepackage{mathptmx}
\usepackage{etoolbox}

\makeatletter
\let\@fnsymbol\@fnsymbol@latex
\@booleanfalse\altaffilletter@sw
\makeatother

\begin{document}

% \preprint{AIP/123-QED}

\title{Macrotransport of active particles in periodic channels and fields: rectification and dispersion}
% Force line breaks with \\
\author{Zhiwei Peng}
\email{zhiwei.peng@ualberta.ca}
\affiliation{Department of Chemical and Materials Engineering, University of Alberta, Edmonton, Alberta T6G 1H9, Canada}
% \author{A. Author}
%  \altaffiliation[Also at ]{Physics Department, XYZ University.}%Lines break automatically or can be forced with \\
% \author{B. Author}%
%  \email{Second.Author@institution.edu.}
% \affiliation{ 
% Authors' institution and/or address%\\This line break forced with \textbackslash\textbackslash
% }%

% \author{C. Author}
%  \homepage{http://www.Second.institution.edu/~Charlie.Author.}
% \affiliation{%
% Second institution and/or address%\\This line break forced% with \\
% }%

\date{\today}% It is always \today, today,

\begin{abstract}
Transport and dispersion of active particles in structured environments such as corrugated channels and porous media are important for the understanding of both natural and engineered active systems. Owing to their continuous self-propulsion, active particles exhibit rectified transport under spatially asymmetric confinement. While progress has been made in experiments and particle-based simulations, a theoretical understanding of the effective  long-time transport dynamics in spatially periodic geometries remains less developed. In this paper, we apply generalized Taylor dispersion theory (GTDT) to analyze the long-time effective  transport dynamics of active Brownian particles (ABPs) in periodic channels and fields. We show that the long-time transport behavior is governed by an effective advection-diffusion equation. The derived macrotransport equations  allow us to characterize the average drift and effective dispersion coefficient. For the case of ABPs subject to a no-flux boundary condition at the channel wall, we show that regardless of activity, the average drift is given by the net diffusive flux along the channel. For ABPs, their activity is the driving mechanism that sustains a density gradient, which ultimately leads to rectified motion along the channel. Our continuum theory is validated against direct Brownian dynamics simulations of the Langevin equations governing the motion of each ABP.
\end{abstract}

\maketitle

\section{\label{sec:intro}Introduction}
Transport of active particles in confined environments are important for the understanding of both natural and engineered active systems. Examples include upstream contamination of motile bacteria\cite{figueroa2020coli,zhou2024ai} and the use of micromotors for cargo delivery\cite{gao2014synthetic,xu2018sperm,de2017micromotor}. The interplay between self-propulsion, geometric confinement, and fluid flows often leads to interesting transport dynamics that are unattainable for passive Brownian systems\cite{romanczuk2012active,bechinger2016active}. Owing to their persistent self-propulsion, active particles accumulate at confining boundaries. When placed under a pressure-driven channel flow, they can migrate against the flow. 

The accumulation of active particles at boundaries depends on the curvature~\cite{yan_brady_2015,smallenburg2015swim,YB2018}. When the obstacle or boundary exhibit spatial asymmetry, active motion can be rectified. For example, particle-based simulations and experiments have shown that swimming bacteria can power gears with asymmetric teeth~\cite{hiratsuka2006microrotary,Angelani,Sokolov,Leonardo,jerez2020dynamics,xu2021rotation}. When the asymmetric boundary is held fixed, active particles will then produce a net flux due to this ratchet effect~\cite{Ghosh2013,Ai2013,ao2014active,ghosh2014giant,ai2016ratchet,RR}. Such rectified transport has been considered for active particles in a spatially periodic channel (see Fig. \ref{fig:schematic} for a schematic). To achieve a non-zero net flux through the channel, the boundary of the channel unit cell must exhibit fore-aft asymmetry. Other geometries such as funnel arrays and periodic porous media with asymmetric pillars have also been considered~\cite{Drocco2012,reichhardt2013active,kantsler2013ciliary,Shelley2017,Tong2018}. In the absence of geometric confinement, one can also obtain rectification with spatially asymmetric ratchet potential fields~\cite{angelani2011active,RR,ai2017transport,nikola2016active}. For a review on ratchet effects in active systems, see \citet{RR}.

Previous works on the rectified transport of active particles through corrugated channels mainly focused on particle-based simulations. In such simulations, the Langevin equations of motion governing the evolution of the stochastic trajectories of active particles are numerically integrated in time. An appropriate ensemble average of the stochastic trajectories allows one to characterize the mean and mean-squared displacements, from which the average speed and dispersion coefficient can be calculated. 

To complement previous simulation results, in this work we develop a statistical mechanical description of the effective long-time longitudinal transport dynamics of active particles in spatially periodic geometries. We restrict our consideration to dilute systems where only single particle dynamics is important. In addition, hydrodynamic interactions between the particle and the channel walls are neglected. With these assumptions, we adopt the active Brownian particle (ABP) model to describe the active dynamics (see Sec. \ref{sec:formulation} for details). Starting from the Smoluchowski equation governing the probability density distribution of an ABP in its position and orientation space, we derive a pair of so-called macrotransport equations that characterize the long-time transport dynamics. More specifically, these equations allow us to calculate the average drift velocity along the channel and the effective longitudinal dispersion coefficient at long times. We consider ABPs in periodic channels, spatially periodic fields, and a combination of geometric channel confinement and external fields.

The macrotransport theory developed here is a generalized Taylor dispersion theory (GTDT) for ABPs in spatially periodic geometries. In this context, one is concerned with the macroscopic dynamics that emerges after ABPs have passed through many copies of the unit cell. The macrotransport equations provide a bridge between the local dynamics and the macroscopic dynamics. The mathematical derivations follow closely those of refs.~\cite{Takatori23,P2024a}. 

For ABPs in a periodic channel in the absence of external flows and fields, we show that the average drift speed ($U^\mathrm{eff}$) along the channel at long times can be given by 
\begin{equation}
\label{eq:intro-drift}
    U^\mathrm{eff} = - D_T \avg{\frac{\partial  \rho_0}{\partial x}}, 
\end{equation}
where $D_T$ is the translational diffusivity of the ABP, $\rho_0$ is the number density field, $x$ is the longitudinal coordinate, and the angle brackets denote the cell average. Equation \eqref{eq:intro-drift} is obtained for ABPs subject to a no-flux boundary condition at the wall. We note that activity (e.g., swim speed) does not appear explicitly in Eq.~\eqref{eq:intro-drift},  which applies to both passive Brownian particles (PBPs) and ABPs. That is, regardless of activity, the drift is given by the net diffusive flux along the channel. For PBPs, the density field is spatially uniform and the drift vanishes. In the case of ABPs, their activity coupled to asymmetric confinement allows the system to maintain a density gradient at steady state, which ultimately leads to rectified transport. As shown by previous work, boundary conditions dictate the macrotransport dynamics.\cite{reichhardt2013active} For active particles that obey an orientation-reflection condition (rule III in ref.\cite{reichhardt2013active}), no rectification is present.

Different from the drift speed $U^\mathrm{eff}$, the swim speed provides a direct contribution to the longitudinal dispersion coefficient---the swim diffusivity modified by asymmetric confinement. In addition, fluctuation in the number density gives another contribution, which is also present for PBPs. In the general case where ABPs are under the influence of both geometric confinement and external fields (such as channel flow), the net swimming motion does contribute directly to the drift. In the main text, we contrast the different contributions to the drift and dispersion coefficient for cases including channel confinement, external fields, orienting fields, and activity landscapes.

This paper is organized as follows. In Sec. \ref{sec:formulation}, we introduce the Smoluchowski formulation and derive the macrotransport equations for ABPs in spatially periodic environments. We first provide a systematic derivation for the case of ABPs in periodic channels without external fields. We then characterize the average drift and the effective longitudinal dispersion, respectively, in Secs. \ref{subsec:average-field-drift} and \ref{subsec:displacement-field-dispersion}. Additional discussions on rectification and dispersion are provided in Sec. \ref{subsec:macrotransport-periodic-channel}. In Sec. \ref{sec:potential-or-fields}, we consider the macrotransport dynamics of ABPs in periodic potentials or fields without geometric confinement. We consider the effect of periodic orienting fields in the absence of channel confinement in Sec. \ref{subsec:orienting-fields}. In Sec. \ref{subsec:combined-channel-fields}, we discuss the combined effects of channel confinement and external fields. Spatially periodic activity landscape is considered in Sec. \ref{subsec:activity-landscape}. In Sec. \ref{sec:asymptotics}, we conduct asymptotic analyses in the limits of weak activity and nearly flat channels. In Sec. \ref{sec:num-examples}, we apply our theory to characterize the macrotransport of ABPs in a corrugated channel and in a one-dimensional substrate potential. We conclude our paper in Sec. \ref{sec:discussion}.

\section{\label{sec:formulation} Problem formulation}
\begin{figure}
\includegraphics[width=3.2in]{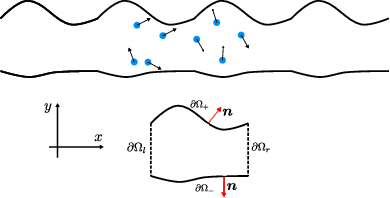}% Here is how to import EPS art
\caption{\label{fig:schematic} Top: Schematic of active Brownian particles in a periodic channel. The particles are treated as point particles in the text. Bottom: Schematic of the unit cell ($\Omega$) and its no-flux ($\partial \Omega_w = \partial \Omega_+ \cup \partial \Omega_-$) and periodic ($\partial \Omega_l \cup \partial \Omega_r$) boundaries. The unit outward normal vector of the channel walls is $\bn$. The unit basis vector in $j$ direction of the Cartesian coordinate system is $\be_j$ for $j=\{x, y,\cdots\}$.  }
\end{figure}

Consider the transport of a dilute suspension of spherical ABPs in a quiescent fluid in a periodic channel (see Fig.~\ref{fig:schematic}). We assume that the radius of the ABPs is negligible compared to the minimum width of the channel, and treat the ABPs as ``point'' particles. In the dilute limit, we further consider the one-particle probability distribution $P(\bX, \bq, t)$ of finding an ABP at position $\bX$ with orientation $\bq$ ($|\bq|=1$) at time $t$. The evolution of $P$ satisfies the Smoluchowski equation: 
\begin{equation}
\label{eq:smol-general}
    \frac{\partial P}{\partial t} + \nabla_X \cdot \bj_T  + \nabla_R\cdot\bj_R=0, 
\end{equation}
where $\nabla_X= \partial /\partial \bX$ and $\nabla_R = \bq \times \partial/\partial \bq  $ are gradient operators in physical and orientation space, respectively. In \eqref{eq:smol-general}, the translational flux $\bj_T$ and rotational flux $\bj_R$ are, respectively, 
\begin{subequations}
\begin{eqnarray}
    \bj_T&=& U_s \bq P - D_T \nabla_X P, \\ 
    \bj_R& =&-D_R \nabla_R P.
\end{eqnarray}
\end{subequations}

The spatial periodicity along the channel ($x$ direction) allows us to relate the global position vector $\bX$ to the position vector $\bx$, defined \emph{locally} in a unit cell, via the relation, 
\begin{equation}
\label{eq:global-local}
    \bX = j L \be_x + \bx, \quad j \in \mathbb{Z}, 
\end{equation}
where $L$ is the spatial periodicity or unit cell length, and $\be_x$ is the unit vector in $x$ direction.  At the top ($\partial \Omega_+$) and bottom channel walls ($\partial \Omega_-$) of the unit cell, the no-flux condition is imposed\cite{ezhilan2015transport}, giving 
\begin{equation}
\label{eq:no-flux-general}
    \bn \cdot\bj_T = 0, \quad \bx \in \partial \Omega_w,
\end{equation}
where $\bn$ is the unit outward normal vector of a boundary, and $\partial\Omega_w=\partial \Omega_+	\cup \partial \Omega_-$ (see figure \ref{fig:schematic} for a schematic of the periodic channel). With \eqref{eq:global-local}, one can then rewrite $P(\bX, \bq, t)$ in terms of the sequence \{$P_j(\bx, \bq, t)$,  $\forall j \in \mathbb{Z}$ \}, where $P_j$ is the probability distribution within the cell $j$.   In other words, to locate the particle in physical space, one can first identify the cell, $j$, in which the particle resides,  and then use the local position $\bx$ within this cell.

% Continuity of the probability density across cell boundaries requires, for all $j \in \mathbb{Z}$, 

% \begin{equation}
% \label{eq:continuity-P}
%         P_j(\bx, \bq, t) = P_{j+1}(\bx - L \be_x , \bq, t), \quad \bx \in \partial \Omega _r. 
% \end{equation}
% Similarly, the continuity of the particle translational flux across cell boundaries implies that, for all $j \in \mathbb{Z}$,  
% \begin{equation}
% \label{eq:continuity-flux}
%     \nabla P_j(\bx, \bq, t) = \nabla P_{j+1}(\bx-L \be_x, \bq, t), \quad \bx \in \partial \Omega_r, 
% \end{equation}
% where $\nabla$ without any subscripts is the local gradient, $\nabla = \partial/\partial \bx$. These continuity relations are generally useful in the context of transport in spatially periodic domains; for example, it is used in the transport of active particles in periodic porous media \cite{Saintillan2019}. 

Taking the zeroth orientational moment of \eqref{eq:smol-general} gives a conservation equation for the number density, 
\begin{subequations}
\label{eq:density-general}
\begin{eqnarray}
        &&\frac{\partial \rho }{\partial t} + \nabla_X \cdot\bj_\rho = 0,  \\ 
        &&\bj_\rho  = U_s \bm{m} - D_T\nabla_X \rho, 
\end{eqnarray}
\end{subequations}
where 
\begin{equation}
    \rho(\bX,t) = \int_{\mathbb{S}} P(\bX, \bq, t) \odiff\bq 
\end{equation}
is the number density, 
\begin{equation}
    \bm{m}(\bX, t) = \int_{\mathbb{S}}  \bq P(\bX, \bq, t)\odiff \bq 
\end{equation}
is the polar order, and $\mathbb{S}$ denotes the unit sphere of orientations in $d$ (e.g., $d=2,3$) spatial dimensions. Unless otherwise specified, the derivation applies to both 2D and 3D. The ``volume'' of the orientation space  is given by $\lvert \mathbb{S}\rvert = \int_{\mathbb{S}} \odiff\bq$. In 2D, $\lvert \mathbb{S}\rvert = 2\pi$; in 3D, $\lvert \mathbb{S}\rvert = 4\pi$.

Because the channel is unbounded along the $x$ direction, it is more convenient to work in Fourier space. To this end, we introduce the semidiscrete Fourier transform \cite{trefethen2000spectral}
\begin{equation}
    \hat{f}(k) = L \sum_{j = -\infty }^\infty e^{- ik j L} f_j, 
\end{equation}
where $k$ is the wavenumber, $i$ ($i^2=-1$) is the imaginary unit, and $f_j$ denotes the function $f$ restricted to the unit cell $j$. We note that the transform is from $j$ to $k$,  and the local spatial coordinate $\bx$ is unchanged. To capture the long-time macrotransport dynamics, we write the derivative $\partial/(\partial \bX)$ as $\partial/(\partial \bx)+ \be_x \partial/[\partial(j L) ]$. At the macroscopic scale, the discrete cell indices can be effectively treated as a continuum. In Fourier space, the Smoluchowski equation \eqref{eq:smol-general} becomes 
\begin{subequations}
\label{eq:smol-general-fourier}
    \begin{equation}
    \frac{\partial \hat{P}}{\partial t} + \left( ik\be_x + \nabla \right)\cdot \hat{\bj}_T + \nabla_R\cdot\hat{\bj}_R=0, 
\end{equation}
\begin{equation}
    \hat{\bj}_T = U_s \bq \hat{P}  - D_T \left( ik\be_x + \nabla \right) \hat{P}, 
\end{equation}
\begin{equation}
    \hat{\bj}_R = -D_R \nabla_R \hat{P},
\end{equation}
\end{subequations}
where $\hat{P} = \hat{P}(k, \bx, \bq, t)$, $\nabla = \partial /\partial \bx$. We note that the sequence of functions $P_j$ reduce to a single function $\hat{P}$ in Fourier space. That is, the dependence on the cell index in real space transforms to a dependence on the wavenumber $k$ in Fourier space.  At long times, we impose periodic boundary conditions on the Fourier-transformed fields~\cite{Takatori23,P2024a}: 
\begin{subequations}
\label{eq:pbc-fourier}
    \begin{equation}
        \hat{P}(\bx, \bq, t) =\hat{P}(\bx - L \be_x, \bq, t), \quad \quad \bx \in \partial \Omega _r, 
    \end{equation}
    \begin{equation}
        \nabla \hat{P}(\bx, \bq, t) = \nabla \hat{P}(\bx-L \be_x, \bq, t), \quad \quad \bx \in \partial \Omega _r.
    \end{equation}
\end{subequations}

Similarly, the number density equation \eqref{eq:density-general} in Fourier space is given by 
\begin{subequations}
\label{eq:density-general-fourier}
\begin{eqnarray}
        &&\frac{\partial \hat{\rho} }{\partial t} + \left( ik\be_x + \nabla \right) \cdot\hat{\bj}_\rho = 0,  \\ 
        &&\hat{\bj}_\rho  = U_s \hat{\bm{m}} - D_T\left( ik\be_x + \nabla \right) \hat{\rho}, 
\end{eqnarray}
\end{subequations}
Averaging \eqref{eq:density-general-fourier} over the unit cell $\Omega$, we have 
\begin{subequations}
\label{eq:avg-n}
    \begin{equation}
    \frac{\partial \avg{\hat{\rho}}}{\partial t} + i k \be_x\cdot \avg{\hat{\bj}_\rho} =0, 
\end{equation}
\begin{equation}
    \avg{\hat{\bj}_\rho}= U_s \avg{\hat{\bm{m}}} - D_T \left( ik \be_x \avg{\hat{\rho}}  
 + \avg{\nabla \hat{\rho}}\right),
\end{equation}
\end{subequations}
where $\avg{(\cdot)} = (1/|\Omega|)\int_\Omega (\cdot) \odiff\bx $, and $|\Omega|$ is the volume (3D) or area (2D) of the unit cell. We note that  $\avg{\hat{\rho}} $ is a function of $k$ and $t$, $\avg{\hat{\rho}} =\avg{\hat{\rho}} (k,t)$. In deriving \eqref{eq:avg-n}, we have applied the divergence theorem, the no-flux boundary conditions on the top and bottom walls, and the periodic condition \eqref{eq:pbc-fourier} on the left and right edges of the unit cell (see Fig. \ref{fig:schematic}). To relate $\hat{P}$ to $\avg{\hat{\rho}}$, we define the ``structure'' function $\hat{G}$ such that 
\begin{equation}\label{eq:G-def}
    \hat{P}(k, \bx, \bq, t) = \avg{\hat{\rho}}(k,t) \; \hat{G}(k, \bx, \bq, t).
\end{equation}
By definition, $\hat{G}$ satisfies the normalization, 
\begin{equation}
\label{eq:G-norm}
    \frac{1}{|\Omega|}\int_\Omega \odiff \bx \int_{\mathbb{S}} \hat{G} \odiff\bq =1.  
\end{equation}

To derive an effective advection-diffusion equation for $\avg{\hat{\rho}}$, we first Taylor expand 
 $\hat{G}$ about $k=0$, giving 
\begin{equation}
\label{eq:small-k-G}
    \hat{G}(k, \bx, \bq,t) = g(\bx, \bq, t) + i\,k\, b(\bx, \bq, t) + O(k^2),
\end{equation}
where $g$ is the zero-wavenumber probability distribution (or the average field) and $b$ is the displacement field \cite{morris_brady_1996,zia_brady_2010,Takatori14,Burkholder17,PB2020, Takatori23}. From \eqref{eq:G-def} and \eqref{eq:small-k-G}, we have  
\begin{subequations}
\label{eq:small-k-n-m}
    \begin{eqnarray}
         \frac{\hat{\rho}}{\avg{\hat{\rho}}} &=&  \rho_0 + i\, k\, \rho_1 +O(k^2) , \\ 
         \frac{\hat{\bm{m}}} {\avg{\hat{\rho}}} &= &  \bm{m}_0 + i\, k\, \bm{m}_1 +O(k^2) , 
    \end{eqnarray}
\end{subequations}
where 
\begin{equation}
    \rho_0 = \int_{\mathbb{S}} g\, \odiff \bq, \quad\text{and} \quad \rho_1 =\int_{\mathbb{S}} b\, \odiff \bq,  
\end{equation}
and $\bm{m}_0$ and $\bm{m}_1$ are similarly defined. We note that an expansion similar to \eqref{eq:small-k-n-m} can be written for any orientational moment of $\hat{P}$. Inserting \eqref{eq:small-k-n-m} into \eqref{eq:avg-n}, one can obtain an effective long-time advection-diffusion equation for $\avg{\hat{\rho}}$ in Fourier space: 
\begin{equation}
\label{eq:effective-1d-eq}
    \frac{\partial  \avg{\hat{\rho}}}{\partial t} + ik\, U^\mathrm{eff}  \avg{\hat{\rho}} + k^2 \,D^\mathrm{eff}  \avg{\hat{\rho}}=0, 
\end{equation}
where the average drift and effective longitudinal dispersion coefficient are, respectively, given by 
\begin{subequations}
    \begin{eqnarray}
    \label{eq:Ueff-eq-general}
        U^\mathrm{eff}&=& U_s  \be_x \cdot \avg{\bm{m}_0} - D_T \be_x \cdot \avg{\nabla \rho_0}, \\ 
        \label{eq:Deff-eq-general}
        D^\mathrm{eff}&=& D_T - U_s \be_x\cdot \avg{\bm{m}_1} + D_T \be_x\cdot\avg{\nabla \rho_1}.
    \end{eqnarray}
\end{subequations}
In writing \eqref{eq:effective-1d-eq}, it is understood that higher-order terms in $k$ are truncated since they do not contribute to either the drift or dispersion. We emphasize that \eqref{eq:effective-1d-eq} does not capture the short-time or transient dynamics. In the context of GTDT, the transport regime prior to the asymptotically long-time regime is often refered to as the pre-asymptotic regime.\cite{taghizadeh2020preasymptotic,guan2023pre}

Subtracting \eqref{eq:avg-n} multiplied by $\hat{G}$ from \eqref{eq:smol-general-fourier}, we obtain 
\begin{eqnarray}
\label{eq:G-eq}
    &&\frac{\partial \hat{G}}{\partial t} + ik\be_x\cdot \left[ U_s\left(\bq - \frac{\avg{\hat{\bm{m}}}}{\avg{\hat{\rho}}}\right)\hat{G}- D_T \left( \nabla \hat{G} - \frac{\avg{\nabla \hat{\rho}}}{\avg{\hat{\rho}}}\hat{G} \right)\right]\nonumber \\ 
    && + \nabla\cdot\left( U_s\bq \hat{G} - D_T ik \be_x \hat{G} - D_T \nabla \hat{G}\right)-D_R \nabla_R^2 \hat{G}=0.
\end{eqnarray}
The no-flux condition \eqref{eq:no-flux-general} is written as 
\begin{equation}
    \bn\cdot\left[ U_s \bq \hat{G} - D_T(ik\be_x + \nabla)\hat{G} \right]=0, \quad \bx \in \partial \Omega_w.
\end{equation}
In $x$ direction, $\hat{G}$ satisfies the periodic conditions, see \eqref{eq:pbc-fourier}. 

\subsection{The average field and the average drift}
\label{subsec:average-field-drift}
Inserting the expansion \eqref{eq:small-k-G} into \eqref{eq:G-eq}, one obtains at $O(1)$ the governing equation 
\begin{equation}
\label{eq:g-eq}
    \frac{\partial g}{\partial t} + \nabla\cdot\left(U_s\bq g - D_T \nabla g\right) - D_R \nabla_R^2 g=0.
\end{equation}
The no-flux condition is given by 
\begin{equation}
\label{eq:no-flux-g}
    \bn\cdot\left(U_s\bq g - D_T \nabla g\right)=0, \quad \bx \in \partial \Omega_w,
\end{equation}
and the periodic conditions \eqref{eq:pbc-fourier} require 
\begin{subequations}
\label{eq:pbc-g}
    \begin{equation}
        g(\bx, \bq, t) =g(\bx - L \be_x, \bq, t), \quad \quad \bx \in \partial \Omega _r, 
    \end{equation}
    \begin{equation}
        \nabla g(\bx, \bq, t) = \nabla g(\bx-L \be_x, \bq, t), \quad \quad \bx \in \partial \Omega _r.
    \end{equation}
\end{subequations}
Lastly, from \eqref{eq:G-norm}, we have the conservation of particles given by 
\begin{equation}
\label{eq:g-norm}
    \frac{1}{|\Omega|}\int_\Omega \odiff \bx \int_{\mathbb{S}} g\, \odiff\bq =1.
\end{equation}
Equations \eqref{eq:g-eq}--\eqref{eq:g-norm} fully specify the problem at leading order, i.e., $O(1)$. With $g$, one can then evaluate the orientational moments $\rho_0$ and $\bm{m}_0$, from which the average drift can be determined using \eqref{eq:Ueff-eq-general}.

The zeroth moment of $g$, or the number density, is governed by 
\begin{subequations}
\label{eq:n0-eq}
    \begin{equation}
    \label{eq:n0-eq-PDE}
        \frac{\partial \rho_0}{\partial t} + \nabla\cdot\left( U_s \bm{m}_0 - D_T \nabla \rho_0 \right)=0, 
    \end{equation}
    \begin{equation}
    \label{eq:n0-eq-noflux}
        \bn\cdot\left( U_s \bm{m}_0 - D_T \nabla \rho_0\right)=0, \quad \bx \in \partial \Omega_w, 
    \end{equation}
\end{subequations}
and the periodic conditions follow from \eqref{eq:pbc-g}.  The first moment of $g$ (the polar order $\bm{m}_0$) satisfies the equation 
\begin{eqnarray}
     \frac{\partial \bm{m}_0}{\partial t} &+& \nabla\cdot\left[U_s \left(\bm{Q}_0+ \frac{ \rho_0\bI}{d}\right)- D_T \nabla \bm{m}_0 \right] \nonumber \\
      &+& (d-1)D_R \bm{m}_0=\bm{0},
\end{eqnarray}
where 
\begin{equation}
    \bQ_0 = \int_{\mathbb{S}}  (\bq\bq - \bI/d)g  \odiff\bq
\end{equation}
 is the nematic order and $\bI$ is the identity tensor. Using the no-flux and periodic conditions, one can show that 
\begin{equation}
\label{eq:m0-avg-eq}
    \frac{\partial\avg{\bm{m}_0} }{\partial t} + (d-1) D_R \avg{\bm{m}_0}=0.
\end{equation}
Any initial polar order would decay to zero exponentially fast due to rotary diffusion \footnote{The average polar order would not vanish in the presence of external orienting fields.}:  $\avg{\bm{m}_0} \to \bm{0}$ as $t\to \infty$. From this, we see that the self-advective (or swimming) contribution to the average drift in \eqref{eq:Ueff-eq-general} vanishes at long times. Using the divergence theorem and periodic conditions, we have 
\begin{equation}
   \label{eq:Ueff-diffusive}
    U^\mathrm{eff} = - D_T \be_x \cdot \avg{\nabla \rho_0} =-\frac{D_T}{|\Omega|} \be_x \cdot \int_{\partial \Omega_w } \bn \,  \rho_0\, \odiff S, 
\end{equation}
where $\odiff S$ is the surface element in 3D and line element in 2D, and $\bn$ is the unit normal vector of the boundary that points out of the domain $\Omega$. 

In deriving \eqref{eq:Ueff-diffusive}, the swim speed is taken to be a constant. If the swim speed depends on the local position $\bx$, $U_s = U_s(\bx)$, the average drift should be written as $U^\mathrm{eff} = \be_x\cdot\avg{U_s \bm{m}_0} - D_T\be_x\cdot\avg{\nabla \rho_0}$. In contrast to \eqref{eq:Ueff-eq-general},  the swim speed needs to be included in the averaging procedure. At long times, while the net polar order still vanishes, the swim flux $\avg{U_s\bm{m}_0}$ does not necessarily vanish.  With a position-dependent swim speed, the average drift has contributions from both the net swim flux and the diffusive flux.

Equation \eqref{eq:Ueff-diffusive} can be deduced from the momentum balance of the particle phase in a unit cell. To see this, consider the micromechanical equation of motion of an active particle \cite{YB2015}, which is given by 
\begin{equation}
    \bm{0} = -\zeta \bU +\bF^\mathrm{swim} + \bF^B  + \bF^\mathrm{ext} + \bF^P, 
\end{equation}
where the swimming motion is driven by the swim force, $\bF^\mathrm{swim} = \zeta U_s \bq$, $\zeta$ is the drag coefficient (recall that hydrodynamic interactions are absent), $\bF^B$ is the fluctuating Brownian force, $\bF^\mathrm{ext}$ is the total external force on the particle, and $\bF^P$ are the particle-particle interactive or collisional force, and $\bU$ is the instantaneous velocity of the particle. In the dilute limit, summing over all particles lead to the global force balance $\zeta \avg{\bU} = \bF^w$, where $\bF^w$ is the force that the top and bottom walls exert on the particles. We note that in writing this average relation we have used the fact that the net polar order vanishes (see \eqref{eq:m0-avg-eq}). The force due to the wall is given by\cite{Brady1993, yan_brady_2015}
\begin{equation}
\label{eq:force-wall2particles}
  \bF^w = - \int_{\partial \Omega_w} \rho k_B T \bn \odiff S,  
\end{equation}
where we have a minus sign because the integral (without the minus sign) is the net force the particles exert on the wall, $k_BT$ is the thermal energy, and $\rho k_BT$ is the osmotic pressure \cite{Brady1993}. Using the Stokes-Einstein-Sutherland relation, $\zeta D_T=k_BT$, we arrive at the statement that $U^\mathrm{eff} = \avg{\bU}\cdot\be_x = F_x^w/\zeta = -D_T \be_x\cdot \int_{\partial \Omega_w}\bn\, \rho\, \odiff S $. Lastly, $\rho$ here is dimensional and can be related to $\rho_0$ via $\rho= \rho_0/|\Omega|$. As a result, one obtains precisely \eqref{eq:Ueff-diffusive} from the momentum balance. The particles exhibit net motion simply to obey the conservation of momentum. Put differently, the channel walls can direct active particles and achieve autonomous directed motion from otherwise random active motion observed in free space. This is the so-called rectification or ratcheting of active Brownian motion. 

It is clear that \eqref{eq:Ueff-diffusive} applies to both active and passive Brownian motion, just like \eqref{eq:force-wall2particles} for the wall force. One may be tempted to think that the average polar order multiplied by the swim speed $U_s$ contributes \emph{directly} to the drift velocity. As seen from the discussion leading from \eqref{eq:Ueff-eq-general} to \eqref{eq:Ueff-diffusive}, this is not the case; the net polar order vanishes. The only contribution--- regardless of activity---is from the diffusive motion driven by the net density gradient. For passive Brownian particles, the density is uniform and we do not observe an average drift. When activity is present, a density gradient can be generated; this is the well-known phenomenon of boundary accumulation of active particles \cite{Rothschild63,Li09,Li11,Costanzo_2012,Elgeti_2013,YB2015,yan_brady_2015}. In a channel with a fore-aft asymmetric boundary, the accumulation or density varies along the channel as a result of curvature variation \cite{YB2018}. While the net polar order vanishes, it contributes \emph{indirectly} to the drift by driving the density out of equilibrium (see \eqref{eq:n0-eq}) and maintaining a steady-state density gradient. Ultimately, this process gives rise to net drift or rectified motion. For passive Brownian particles, this intrinsic driving force is absent, and rectification is often achieved by applying external forces that bias and direct particles.\cite{Hanggi1994,hanggi1996brownian,Hanggi2009} 

Rectification allows us to harness the otherwise random---albeit active---motion of active particles to generate directed motion or perform useful work. While our focus here is on the autonomous macrotransport of active particles in channels, one can also achieve directed motion of a water permeable container that encapsulates active particles \cite{peng_zhou_brady_2022}. In exterior problems,   objects immersed in an active bath are shown to exhibit spontaneous and directed rotation \cite{Angelani,Leonardo,Sokolov,Ray}. For a review on ratchet effects in active matter, see \citet{RR}.

We notice that \eqref{eq:Ueff-diffusive} was derived by \citet{Yariv14}. In their work, the average drift was considered in real (or physical) space instead of in Fourier space. Indeed, one may drop the dependence on the global coordinate $jL$ from the outset and assume periodic conditions on $P$, which leads directly to \eqref{eq:g-eq}. The average drift can then be obtained by averaging the translational flux for the density given in \eqref{eq:n0-eq}. This real-space approach is the leading-order moment in the so-called Aris's method of moments in the context of Taylor dispersion \cite{Aris1956, barton_1983, frankel_brenner_1989,brenner1993macrotransport,Saintillan2019,jiang_chen_2019}.  One can continue to higher order moments and obtain, for example, the effective longitudinal dispersion coefficient; this was not considered in \citet{Yariv14}. We note that Aris's method of moments and the Fourier approach lead to the same drift and dispersion coefficients in general macrotransport processes. We find it more intuitive to work in Fourier space since the wavenumber $k$ provides a natural parameter for ordering the drift and dispersion coefficient (see, for example, equation \eqref{eq:effective-1d-eq}).

% \begin{figure}
% \includegraphics{fig_1}% Here is how to import EPS art
% \caption{\label{fig:epsart} A figure caption. The figure captions are
% automatically numbered.}
% \end{figure}
% is small enough to fit in a single column, while
% Fig.~\ref{fig:wide}%
% \begin{figure*}
% \includegraphics{fig_2}% Here is how to import EPS art
% \caption{\label{fig:wide}Use the \texttt{figure*} environment to get a wide
% figure, spanning the page in \texttt{twocolumn} formatting.}
% \end{figure*}
\subsection{The displacement field and the effective dispersion}
\label{subsec:displacement-field-dispersion}
Equation \eqref{eq:G-eq} at $O(k)$ gives 
\begin{eqnarray}
\label{eq:d-eq}
      && \frac{\partial b}{\partial t} + \nabla\cdot\left(U_s\bq b - D_T \nabla b\right) - D_R \nabla_R^2 b\nonumber \\
      &&= 2 D_T \frac{\partial g}{\partial x} + \left( U^\mathrm{eff} - U_s q_x\right)  g, 
\end{eqnarray}
where $q_x = \bq\cdot\be_x$. The boundary condition at the top and bottom of the unit cell is given by 
\begin{equation}
\label{eq:no-flux-d}
    \bn\cdot\left(U_s\bq b - D_T \nabla b\right)=\bn\cdot\be_x\, D_T\, g, \quad \bx \in \partial \Omega_w,
\end{equation}
In $x$ direction, $b$ obeys the periodic conditions similar to \eqref{eq:pbc-g}. The conservation condition \eqref{eq:G-norm} implies that 
\begin{equation}
\label{eq:d-norm}
    \int_\Omega \odiff \bx \int_{\mathbb{S}} b\,  \odiff\bq =0.
\end{equation}
Notice that the operators on the left-hand side of \eqref{eq:d-eq} is the same as \eqref{eq:g-eq}, and the displacement field is driven by the $g$ field. More specifically, it is driven by a diffusive flux plus the deviation of the swimming motion from the mean drift. We also note that the boundary condition \eqref{eq:no-flux-d} is also non-homogeneous. 

To determine the effective dispersion coefficient, the number density and polar order of $b$ are needed. Taking the zeroth orientational moment of \eqref{eq:d-eq} gives the equation for $\rho_1$: 
\begin{equation}
\label{eq:n1-eq}
    \frac{\partial \rho_1}{\partial t} + \nabla\cdot\left( U_s \bm{m}_1 - D_T \nabla \rho_1  \right) = 2D_T \frac{\partial \rho_0}{\partial x}+ U^\mathrm{eff}\rho_0 - U_s m_{0,x},
\end{equation}
where $m_{0,x} = \bm{m}_0\cdot\be_x$. The boundary condition on the wall follows from \eqref{eq:no-flux-d}, and is given by 
\begin{equation}
\label{eq:no-flux-n1}
    \bn\cdot\left(  U_s \bm{m}_1 - D_T \nabla \rho_1 \right) = D_T \rho_0 \bn\cdot\be_x, \quad \bx \in \partial \Omega_w.
\end{equation}
The total density at $O(k)$ is zero, $\int_\Omega \rho_1 d\bx =0$. Using \eqref{eq:no-flux-n1} and the periodic condition, one can show that the average of \eqref{eq:n1-eq} leads to the equation $0 = D_T \be_x\cdot\avg{\nabla \rho_0} +U^\mathrm{eff}$, which is consistent with \eqref{eq:Ueff-diffusive}. The governing equation for the polar order is given by 
\begin{eqnarray}
\label{eq:m1-eq}
    &&\frac{\partial \bm{m}_1}{\partial t} + \nabla\cdot\left[ U_s \left( \bQ_1 + \frac{1}{d}\rho_1\bI \right) - D_T \nabla \bm{m}_1  \right] +(d-1)D_R \bm{m}_1\nonumber \\ 
    && = 2D_T \frac{\partial \bm{m}_0}{\partial x}+ U^\mathrm{eff}\bm{m}_0 - U_s \left(\bQ_0 + \frac{1}{d}\rho_0\bI\right)\cdot\be_x. 
\end{eqnarray}
At the top and bottom walls, $\bx \in \partial \Omega_w$, we have 
\begin{equation}
    \bn\cdot\left[  U_s \left( \bQ_1 + \frac{1}{d}\rho_1\bI \right) - D_T \nabla \bm{m}_1 \right] = D_T \bn\cdot\be_x\bm{m}_0.
\end{equation}
The polar order is also periodic in $x$ direction.

Averaging \eqref{eq:m1-eq} in the unit cell at long times gives
\begin{equation}
\label{eq:m1-avg-sol}
    \avg{\bm{m}_1} = - \frac{\ell}{d(d-1)}\be_x  + \frac{\delta^2}{d-1} \bavg{\nabla\cdot\left( \be_x \bm{m}_0\right) },
\end{equation}
where we have used the relations $\avg{\bm{m}_0} = \bm{0}$ (see \eqref{eq:m0-avg-eq}) and $\avg{\bQ_0} = \bm{0}$ \footnote{This can be obtained by an argument similar to \eqref{eq:m0-avg-eq}. }, $\delta=\sqrt{D_T\tau_R}$ is the microscopic length, and $\ell=U_s\tau_R$ is the run (or persistence) length. Inserting \eqref{eq:m1-avg-sol} into \eqref{eq:Deff-eq-general}, we have 
\begin{equation}
\label{eq:D-eff-general-simp}
    D^\mathrm{eff} = D_0^\mathrm{eff} - \frac{U_s\delta^2}{d-1} \be_x\be_x: \bavg{\nabla \bm{m}_0} + D_T \be_x \cdot \bavg{\nabla \rho_1},
\end{equation}
where $D_0^\mathrm{eff} = D_T + D^\mathrm{swim}_0$ is the effective diffusivity of an ABP in free space (or the effective longitudinal diffusivity in a flat channel \cite{PB2020}) and $D_0^\mathrm{swim} = U_s^2\tau_R/[d(d-1)]$ is the swim diffusivity. Notice that $D_0^\mathrm{eff}$ emerges naturally from the above consideration and the effect of the channel confinement appear as additional contributions (or corrections) in \eqref{eq:D-eff-general-simp}. We also remark that $\rho_1$ has units of length, which follows from the expansion \eqref{eq:small-k-G}. 

\subsection{Macrotransport of active particles in periodic channels}
\label{subsec:macrotransport-periodic-channel}
Equations \eqref{eq:Ueff-diffusive} and \eqref{eq:D-eff-general-simp} are the main results of this paper. To calculate the average drift in \eqref{eq:Ueff-diffusive}, one needs to solve for the average field $g$ using \eqref{eq:g-eq}--\eqref{eq:g-norm},  and then take the zeroth orientational moment to obtain $\rho_0$. Similarly, the longitudinal dispersion depends on the solution to the displacement field $b$, which is governed by \eqref{eq:d-eq}--\eqref{eq:d-norm}.  Equations \eqref{eq:Ueff-diffusive}, \eqref{eq:D-eff-general-simp} and the aforementioned governing equations for the average and displacement fields provide a complete description of macrotransport of active particles in periodic channels. We remark that no assumptions regarding the geometry of channel unit cell have been made and no approximations have been made in deriving the macrotransport equations beyond the Smoluchowski equation \eqref{eq:smol-general}. Therefore, these equations apply to generic periodic channels and provides an exact characterization of the macrotransport of active (or passive) particles.

In \eqref{eq:Ueff-diffusive}, we have shown that the average drift results from the net diffusive transport along the channel. For passive particles regardless of the channel geometry, the density gradient vanishes and the average drift is zero. For active particles, if the channel walls are flat and parallel, one can see that the net flux vanishes due to symmetry. As a result, for passive particles, external steady and/or oscillatory driving forces are often employed in order to achieve rectified transport; for active particles, however, fore-aft asymmetric channel confinement alone is able to drive a nonzero average drift. That is, active particles under appropriate confinement exhibit \emph{autonomous} rectification. 

In addition to the average drift or rectified transport, we also derived the long-time effective longitudinal dispersion coefficient, as given by \eqref{eq:D-eff-general-simp}. For passive particles, $U_s \equiv 0$, one can readily show that $g = 1/|\mathbb{S}|$, $\rho_0 = 1$, and $\bm{m}_0 = \bm{0}$. From \eqref{eq:n1-eq} and \eqref{eq:no-flux-n1}, we then have 
\begin{subequations}
\label{eq:rho-1-passive}
    \begin{equation}
        \nabla^2 \rho_1 = 0, 
    \end{equation}
    \begin{equation}
        \bn \cdot \nabla \rho_1 = - \bn\cdot\be_x, \quad \bx \in \partial \Omega_w. 
    \end{equation}
\end{subequations}
From \eqref{eq:D-eff-general-simp}, we obtain for passive particles
\begin{equation}
\label{eq:D-eff-passive}
    \frac{D^\mathrm{eff}_\mathrm{passive}}{D_T} = 1 +   \Bavg{\frac{\partial \rho_1}{\partial x}}.
\end{equation}
In the absence of activity, one only needs to solve the Laplace equation for $\rho_1$ subject to a Neumann boundary condition on $\partial \Omega_w$ and periodicity in $x$. While channel confinement cannot give rise to rectified transport of passive particles, it modifies the dispersion coefficient. As noted earlier, regardless of activity, the average drift is given by the net diffusive transport---activity plays a role by driving the density out of equilibrium. This is not the case for the dispersion coefficient (see \eqref{eq:D-eff-passive} and \eqref{eq:D-eff-general-simp}). Compared to the passive result \eqref{eq:D-eff-passive}, the active result in \eqref{eq:D-eff-general-simp} has an additional contribution due to the fluctuations in the swimming motion or polar order. The fluctuations in swimming motion manifest as a swim diffusivity. Because of \eqref{eq:m1-avg-sol}, this swimming contribution can be given by the undisturbed swim diffusivity and the additional contribution from the confinement.

While our aim was to derive a macrotransport theory for active particles in periodic (nonflat) channels, we can recover the transport problem of active particles in flat channels. To this end, we first note that for flat channels the unit cell reduces to a rectangular domain and the width $L$ along the channel can be arbitrarily chosen. For flat channels, the average and displacement fields do not depend on $x$. The average field equation \eqref{eq:g-eq} reduces to 
\begin{equation}
\label{eq:g-eq-flat}
    \frac{\partial}{\partial y }\left(U_s q_y g - D_T \frac{\partial g}{\partial y} \right) - D_R \nabla_R^2 g=0,
\end{equation}
and the no-flux condition is $U_s q_y g - D_T \frac{\partial g}{\partial y}=0$ at $y = \pm H$. Here, we have assumed that the flat channel has a transverse width of $2H$ with the top and bottom walls, respectively,  located at $y=H$ and $y=-H$. Because the wall normal $\bn = \pm \be_y$, from \eqref{eq:Ueff-diffusive} we see that $U^\mathrm{eff}=0$, as expected. Equation \eqref{eq:g-eq-flat} governs the distribution of active particles confined between two flat plates, which is a well-studied problem. Notably, due to their self-propulsion, active particles accumulate at the wall.  The displacement field equation \eqref{eq:d-eq} can be written as 
\begin{eqnarray}
\label{eq:d-eq-flat}
\frac{\partial }{\partial y}\left(U_sq_y b - D_T \frac{\partial b}{\partial y}\right) - D_R \nabla_R^2 b =   - U_s q_x g.
\end{eqnarray}
The boundary conditions  are given by 
\begin{equation}
    U_sq_y b - D_T \frac{\partial b}{\partial y}=0, \quad y=\pm H. 
\end{equation}
We notice that the boundary conditions are homogeneous (cf. \eqref{eq:no-flux-d}) since $\bn \cdot \be_x =0$. From \eqref{eq:D-eff-general-simp}, it is clear that $D^\mathrm{eff} = D_0^\mathrm{eff}$. For flat channels, we see that the average drift is zero and the longitudinal dispersion equals that in free space \cite{PB2020}. In the presence of Poiseuille flow through flat channels, active particles are shown to exhibit upstream swimming and the longitudinal dispersion coefficient varies non-monotonically as a function of the flow speed \cite{PB2020}. In the presence of such flows, equations \eqref{eq:g-eq-flat} and \eqref{eq:d-eq-flat} need to include the appropriate flow terms; similarly, additional contributions to the drift and dispersion from the background flow need to be considered (see ref. \cite{PB2020}).

 In formulating the macrotransport equations, we took as an example a channel with two separate walls (see Fig.~\ref{fig:schematic}). We note that this is only for concreteness,  and the derived equations apply equally well to a tube with periodically varying cross-section. The only modification is that the no-flux portion of the unit cell boundary ($\partial \Omega_w$) is now a single continuous surface instead of two separate walls.

 Lastly, we note that our formulation can also be extended to the transport of active particles in periodic porous media or substrate.\cite{bhattacharjee2019bacterial,Saintillan2019,Reichhardt21,Kumar2022,reichhardt2023pattern} In \citet{Saintillan2019}, Aris's moment method (real space) is used to study the drift and dispersion of active particles in flows through periodic porous media. We note that in a regular lattice of pillars one needs to take the semidiscrete Fourier transform in all Cartesian directions (see Sec. \ref{sec:potential-or-fields} for an example).  The wall condition is now applied at the interior boundary of the unit cell (the surface of the pillar), and at the outer sides periodic conditions apply. If the pillars are fore-aft asymmetric, rectified transport in porous media can also be achieved \cite{Shelley2017,Tong2018}.

\section{Macrotransport in periodic fields}

\subsection{Periodic potentials or external fields}
\label{sec:potential-or-fields}
In the absence of confining asymmetric channels, directed transport of active particles can be achieved using asymmetric external fields or potentials \cite{Takatori14,RR,Liao2018,zhen2022optimal}. In symmetric potential fields, the dispersion behavior of active particles is also non-trivial~\cite{Takatori23,modica2024soft}. We consider the external force $\bF^\mathrm{ext}$ to be spatially periodic such that $\bF^\mathrm{ext}(\bx+\bL, t) = \bF^\mathrm{ext}(\bx, t)$, where $\bL = \sum_{j=1}^d z_j L_j \be_j$  (To be consistent with our notation, $\be_1=\be_x$, $\be_2=\be_y$, etc) with $L_j$ being the periodicity and $z_j\in \mathbb{Z}$ the cell index in $i$ direction, respectively. We note that the force may be applied directly or derived from a periodic potential. Furthermore, $\bF^\mathrm{ext}$ may also be time periodic with period $T$,  $\bF^\mathrm{ext}(\bx, t+T) =\bF^\mathrm{ext}(\bx,t)$, in which case an additional time average over a period $T$ is needed to calculate the average drift and effective dispersion.  In the Smoluchowski equation \eqref{eq:smol-general}, we only need to include an additional translational flux term due to the external force given by $\bU^\mathrm{ext} g$ with $\bU^\mathrm{ext}= \bF^\mathrm{ext} /\zeta$.  

The Fourier transform is now carried out in all directions. Instead of the scalar wavenumber $k$, we have the wave vector $\bk$ and the displacement field is a vector field. As a result, the small wavenumber expansion is $\hat{G} = g + i\bk \cdot\bb +O(k^2)$, where $k$ is the norm of $\bk$. Because the remaining derivation is similar to that in a periodic channel, in the following we only summarize the main results. For details of the derivation, see the supplementary material.

One can show that the average field is governed by 
\begin{equation}
        \frac{\partial g}{\partial t} + \nabla\cdot\left(U_s\bq g + \bU^\mathrm{ext}g - D_T \nabla g\right) - D_R \nabla_R^2 g=0,
\end{equation}
where $g$ is spatially periodic. The average drift is written as 
\begin{equation}
\label{eq:U-eff-potential}
    \bU^\mathrm{eff} = \bavg{\bU^\mathrm{ext} \rho_0}.
\end{equation}
The average polar order $\avg{\bm{m}_0}$ remains zero at long times in the presence of the external force.  More importantly, we note that the diffusive term $-D_T \avg{\nabla \rho_0}$ vanishes due to spatial periodicity of the fields (cf. \eqref{eq:Ueff-diffusive}). The only contribution to the drift is from the  external driving force, $\bavg{\bU^\mathrm{ext} \rho_0}$. The conservation condition \eqref{eq:g-norm} remains unchanged. 

The displacement field satisfies the equation, 
\begin{eqnarray}
\label{eq:d-eq-Uext}
      && \frac{\partial \bb}{\partial t} + \nabla\cdot\left(U_s\bq \bb +\bU^\mathrm{ext}\bb- D_T \nabla \bb\right) - D_R \nabla_R^2 \bb\nonumber \\
      &&= 2 D_T \nabla g + \left( \bU^\mathrm{eff} - U_s \bq - \bU^\mathrm{ext}\right)  g, 
\end{eqnarray}
and spatial periodicity. Because $\bb$ is a vector field, the ``density'' is a vector field, $\brho_1 = \int_{\mathbb{S}} \bb \, \odiff\bq$, and the polar order is a second-order tensor, $\bm{m}_1 = \int_{\mathbb{S}} \bb \bq \odiff\bq$.  In the presence of the external force, the effective dispersion tensor reads
\begin{equation}
\label{eq:D-eff-Fext}
\bD^\mathrm{eff} = D_T \bI  - U_s\avg{\bm{m}_1} - \avg{\bU^\mathrm{ext}\brho_1},
\end{equation}
where we have used the fact that $\avg{\nabla \brho_1}=0$. Following the procedure outlined in Sec.~\ref{subsec:displacement-field-dispersion} (see supplementary material), one can show that at long times, 
\begin{equation}
    \avg{\bm{m}_1} = - \frac{\ell}{d(d-1)}\bI  - \frac{\tau_R}{d-1}\avg{\bU^\mathrm{ext} \bm{m}_0}. 
\end{equation}
From \eqref{eq:D-eff-Fext}, we then have 
\begin{equation}
    \label{eq:D-eff-Fext-simp}
    \bD^\mathrm{eff} = D_0^\mathrm{eff}\bI + \frac{\ell}{d-1}\avg{\bU^\mathrm{ext} \bm{m}_0}- \avg{\bU^\mathrm{ext} \brho_1}. 
\end{equation}
We note that \eqref{eq:D-eff-Fext-simp} and \eqref{eq:D-eff-Fext} are equivalent. Here, just like the drift, the driving force gives a direct contribution to the dispersion tensor, $- \bavg{\bU^\mathrm{ext} \brho_1}$. Compared to the free diffusion tensor $D_0^\mathrm{eff}\bI$, the external force drives additional density fluctuations, which ultimately affects the dispersion tensor. In contrast to channel confinement, the gradients of $\bm{m}_0$ and $\brho_1$ do not contribute to the dispersion. 

Next, consider the simple case in which the external force is only a function of $x$ and time $t$, $\bU^\mathrm{ext} =\bU^\mathrm{ext}(x,t)$. That is, the force is invariant in the transverse directions. In this case, one can integrate out the transverse coordinates to obtain 
\begin{equation}
     \frac{\partial g}{\partial t} + \frac{\partial}{\partial x}\left(U_s\,q_x\, g + U_x^\mathrm{ext}g - D_T \frac{\partial g }{\partial x} \right) - D_R \nabla_R^2 g=0,   
\end{equation}
where $g = g(x,\bq, t) = g(x+L, \bq, t)$. As a result, the average drift in $x$ direction becomes 
\begin{equation}
    U_x^\mathrm{eff} = \bavg{U_x^\mathrm{ext} \rho_0},
\end{equation}
where the unit cell is 1D, $\Omega = \{x\,\rvert\, 0 \leq x \leq L\}$, and $\avg{\cdot} =(1/L) \int_0^L (\cdot) dx$. The average drift in the transverse direction is 
\begin{align}
    \bU^\mathrm{eff}_\perp = (\bI - \be_x\be_x)\cdot \bU^\mathrm{eff} = \bavg{\bU_\perp^\mathrm{ext} \rho_0}.
\end{align}
To understand this formula, consider a periodic driving force that gives $\bU^\mathrm{ext} = U_\parallel(x,t)\be_x + U_\perp \be_y$, where $U_\parallel$ is $L$-periodic and $U_\perp$ is constant in space.  With this force, we have $\bU^\mathrm{eff}_\perp = U_\perp \be_y \avg{\rho_0} = U_\perp \be_y $. That is, the particles have a constant speed of translation in $y$. This motion, however, has no influence on the transport in the $x$ direction, as expected.  

% The displacement equation reduces to 
% \begin{subequations}
% \begin{eqnarray}
% \label{eq:bx-1d}
%       && \frac{\partial b_x}{\partial t} + \frac{\partial}{\partial x}\left(U_s \,q_x\, b_x +U_x^\mathrm{ext}b_x- D_T \frac{\partial b_x}{\partial x}\right) - D_R \nabla_R^2 b_x\nonumber \\
%       &&= 2 D_T \frac{\partial g}{\partial x} + \left( U_x^\mathrm{eff} - U_s q_x - U_x^\mathrm{ext}\right)  g, 
% \end{eqnarray} 
% \begin{eqnarray}
% \label{eq:b-perp-1d}
%       && \frac{\partial \bb_\perp}{\partial t} + \frac{\partial}{\partial x}\left(U_s \,q_x\, \bb_\perp +U_x^\mathrm{ext}\bb_\perp- D_T \frac{\partial \bb_\perp}{\partial x}\right) - D_R \nabla_R^2 b_x\nonumber \\
%       &&= 2 D_T \frac{\partial g}{\partial x} + \left( U_x^\mathrm{eff} - U_s q_x - U_x^\mathrm{ext}\right)  g, 
% \end{eqnarray} 
% \end{subequations}

\subsection{Periodic orienting fields}
\label{subsec:orienting-fields}
In an external orienting field without geometric confinement, the active particle acquires an angular velocity $\bomega$, which we assume to be spatially periodic and depends on the particle orientation $\bq$, i.e., $\bomega = \bomega(\bx, \bq)$. In the Smoluchowski equation, the rotational flux is written as $ \bj_R = \bomega(\bx)P-D_R \nabla_R P$. For simplicity, we consider $\bomega = \omega_0 \bq\times \hat{\bH}$, where  $\hat{\bH}$ is a unit vector in the field direction and $\omega_0$ is the amplitude \cite{Takatori14,Dulaney20,Shaik2023}. To achieve a spatially periodic $\bomega$, one may take the amplitude $\omega_0$ to be spatially periodic while $\hat{\bH}$ is constant. We note that other forms of $\omega$ may be considered depending on the context \cite{Takatori14,Yan_2018,Shaik2023}. 

Following previous sections, it is straightforward to obtain the average field equation as 
\begin{eqnarray}
 \label{eq:g-eq-torque}
    \frac{\partial g}{\partial t} &+& \nabla\cdot\left(U_s\bq g - D_T \nabla g\right) \nonumber \\ 
    &+&\nabla_R\cdot\left( \omega_0 \bq \times\hat{\bH}g - D_R \nabla_R g \right) =0.
\end{eqnarray}
The polar order of $g$ (recall that $\bm{m}_0 = \int_{\mathbb{S}}  g \bq\odiff\bq$) satisfies the equation, 
\begin{eqnarray}
     &&\frac{\partial \bm{m}_0}{\partial t} + \nabla\cdot\left[U_s \left(\bm{Q}_0+ \frac{ \rho_0\bI}{d}\right)- D_T \nabla \bm{m}_0 \right] \nonumber \\
      &&+ (d-1)D_R \bm{m}_0+  \omega_0 \left( \bQ_0\cdot\hat{\bH} - \frac{d-1}{d}\rho_0 \hat{\bH}\right)=\bm{0}.
\end{eqnarray}
Averaging this equation over the unit cell, we obtain 
\begin{eqnarray}
    \frac{\partial \avg{\bm{m}_0}}{\partial t} &+& (d-1)D_R \avg{\bm{m}_0} \nonumber \\ 
    &+& \Bavg{\omega_0 \left( \bQ_0\cdot\hat{\bH} - \frac{d-1}{d}\rho_0 \hat{\bH}\right)}=\bm{0}. 
\end{eqnarray}
Due to the external orienting field, the cell-averaged polar order does not necessarily vanish. As a result, the average drift is given by 
\begin{equation}
    \bU^\mathrm{eff}= U_s  \avg{\bm{m}_0}.
\end{equation}
In this case, the net polar order is the only contribution to  the average drift velocity. 

The displacement field is governed by 
\begin{eqnarray}
\label{eq:d-eq-torque}
      && \frac{\partial \bb}{\partial t} + \nabla\cdot\left(U_s\bq \bb - D_T \nabla \bb\right) + \nabla_R\cdot \left(\omega_0 \bq\times \hat{\bH}\bb - D_R \nabla_R \bb \right)\nonumber \\
      &&= 2 D_T \nabla g + \left( \bU^\mathrm{eff} - U_s \bq\right)  g, 
\end{eqnarray}
and the dispersion tensor is 
\begin{equation}
    \bD^\mathrm{eff}= D_T \bI  - U_s  \avg{\bm{m}_1}.
\end{equation}
We notice that fluctuation in the polar order gives a contribution to the diffusivity in addition to $D_T$. This is the swim diffusivity in the presence of a spatially periodic orienting field. For the case where both the field direction $\hat{\bH}$ and the amplitude $\omega_0$ are constant, the swim diffusivity has been considered. \cite{Takatori14} 

\subsection{Combined effects of fields and channel confinement}
\label{subsec:combined-channel-fields}
We now derive the macrotransport equations with  both external fields and channel confinement. The translational flux that enters the Smoluchowski equation \eqref{eq:smol-general} can be written as 
\begin{equation}
\label{eq:jT-all-effects}
    \bj_T = U_s\bq P + \bU^\mathrm{ext}(\bx) P - D_T \nabla_X P,
\end{equation}
where we recall that $\bU^\mathrm{ext}$ represents the velocity induced by an external field as discussed in  section \ref{sec:potential-or-fields}. The rotational flux is given by  
\begin{equation}
\label{eq:jR-all-effects}
    \bj_R = \bomega(\bx, \bq) P -D_R \nabla_R P,  
\end{equation}
where  the field-induced angular velocity $\bomega$ can depend on both the local position $\bx$ and the orientation $\bq$. We note that \eqref{eq:jT-all-effects} and \eqref{eq:jR-all-effects} can be applied to the transport of ABPs in a periodic flow through a periodic corrugated channel, in which case $\bU^\mathrm{ext}$ is the fluid velocity and $\bomega$ is the particle angular velocity (not equal to the vorticity) due to the flow (see Ref. \cite{zhou2024ai}).

Following the procedure given in previous sections, it is straightforward to show that 
\begin{subequations}
\label{eq:U-D-all-effect}
    \begin{align}
        &U^\mathrm{eff} = U_s \be_x \cdot\avg{\bm{m}_0} + \be_x\cdot\bavg{\bU^\mathrm{ext} \rho_0} - D_T \be_x \cdot\bavg{\nabla\rho_0},\\
        &D^\mathrm{eff}= D_T - U_s \be_x\cdot \avg{\bm{m}_1}- \bavg{U_x^\mathrm{ext} \rho_1} + D_T \be_x\cdot\avg{\nabla \rho_1}. 
    \end{align}
\end{subequations}
For a flat channel, the terms involving $\nabla \rho_0$ and $\nabla \rho_1$ in the drift and dispersion coefficients vanish. Furthermore, if the external fields are due to a Poiseuille flow, equation \eqref{eq:U-D-all-effect} reduces to those given in \citet{PB2020} (see also section \ref{subsec:macrotransport-periodic-channel}). 

The average field is governed by
\begin{eqnarray}
   \label{eq:g-eq-all-effects}
   \frac{\partial g}{\partial t}  &+& \nabla\cdot\left(U_s\bq g  + \bU^\mathrm{ext} g - D_T \nabla g\right)\nonumber \\ 
    &+& \nabla_R \cdot\left( \bomega g - D_R \nabla_R g\right)=0.
\end{eqnarray}
At the no-flux boundaries, $\bx \in \partial \Omega_w$, we have
\begin{equation}
   \bn\cdot \left(U_s\bq g  + \bU^\mathrm{ext} g - D_T \nabla g\right)=0. 
\end{equation}
The normalization and periodic conditions remain the same as those given in section \ref{subsec:average-field-drift}. Similarly, one can show that 
\begin{eqnarray}
\label{eq:d-eq-all-effects}
      && \frac{\partial b}{\partial t} + \nabla\cdot\left(U_s\bq b + \bU^\mathrm{ext} b- D_T \nabla b\right) + \nabla_R \cdot\left( \bomega b - D_R \nabla_R b\right)\nonumber\\ 
      &&= 2 D_T \frac{\partial g}{\partial x} + \left( U^\mathrm{eff} - U_s q_x- U_x^\mathrm{ext}\right)  g, 
\end{eqnarray}
and 
\begin{equation}
\label{eq:no-flux-d-all-effects}
    \bn\cdot\left(U_s\bq b  + \bU^\mathrm{ext} b - D_T \nabla b\right)=\bn\cdot\be_x\, D_T\, g, \quad \bx \in \partial \Omega_w. 
\end{equation}
The normalization and periodic conditions for $b$ are given in section \ref{subsec:displacement-field-dispersion}.

\subsection{Spatially periodic activity landscape}
\label{subsec:activity-landscape}
Active particles in an activity landscape are shown to exhibit density and polarization patterns~\cite{Schinitzer93,Tailleur08,Row2020,stenhammar2016light,soker2021activity,Auschra,Wysocki_2022,Part24}. Tuning the spatial-temporal variation of the swim speed of active particles may allow one to control the system behavior~\cite{cates2015motility,arlt2018painting,frangipane2018dynamic}. In the absence of channel confinement, we now consider the case in which the active particles are subject to a one-dimensional unbounded and spatially periodic activity landscape. That is, the scalar swim speed is a function of $x$ with a periodicity of $L$, $U_s(x+L)=U_s(x)$. 

In the 1D activity landscape, one can show that at long times the effective transport coefficients in the $x$ direction are 
\begin{equation}
\label{eq:activity-landscape-coeffs}
      U^\mathrm{eff}=  \be_x\cdot \avg{U_s \bm{m}_0}, \quad\mathrm{and}\quad D^\mathrm{eff} = D_T - \be_x\cdot\avg{U_s \bm{m}_1},
\end{equation}
where the unit cell is $\Omega = \{x \,\lvert\, 0 \leq x \leq L\}$, and the cell-averaging is $\avg{\cdot} = (1/L)\int_0^L (\cdot)\odiff x$. The average field is governed by 
\begin{equation}
    \frac{\partial g}{\partial t} + \frac{\partial}{\partial x}\left( U_s q_x g - D_T \frac{\partial g}{\partial x}\right)- D_R \nabla_R^2 g=0, 
\end{equation}
where $g$ is spatially periodic and satisfies the normalization $\avg{n_0}=1$ with $n_0 = \int_{\mathbb{S}} g \odiff \bq $. The displacement field is governed by 
\begin{align}
     \frac{\partial b}{\partial t} +\frac{\partial }{\partial x}\left(U_s q_x b -D_T\frac{\partial b}{\partial x} \right)- D_R \nabla_R^2 b  =& \left(U^\mathrm{eff} - U_s q_x \right) g \nonumber \\ 
     & + 2 D_T \frac{\partial g}{\partial x},   
\end{align}
where $\avg{\int_\mathbb{S} b \odiff \bq }=0$.

We note that the term $-\be_x \cdot \avg{U_s \bm{m}_1}$ in \eqref{eq:activity-landscape-coeffs} is the swim diffusivity in an activity landscape. For ABPs with a constant speed, at long times $g$ and $b$ are spatially invariant. Supposing that the orientation space is in 2D, we then have $g=1/(2\pi)$ and $b = - \ell/(2\pi)\cos\theta$, where we have taken $\bq = \cos\theta \be_x +\sin\theta\be_y$. From this, we obtain $D^\mathrm{eff} = D_T + U_s \ell/2=D_T +U_s^2\tau_R/2$, which is the long-time diffusivity of an ABP in unbounded 2D space with a constant speed $U_s$. We further note that the macrotransport equations in a doubly or triply periodic activity landscape can be similarly obtained.

\section{Asymptotic analysis}
\label{sec:asymptotics}

\subsection{Weak swimming}
\label{subsec:weak-swim}
For convenience, we scale lengths by the spatial periodicity $L$ of the channel and time by the diffusive timescale $\tau_D=L^2/D_T$. At steady state, the non-dimensional equation for $g$ is
\begin{equation}
\label{eq:g-non-dim}
\nabla\cdot\left(Pe \,\bq\, g - \nabla g\right) - \gamma^2 \nabla_R^2 g=0,
\end{equation}
where we have defined the swimming P\'eclet number 
\begin{equation}
    Pe = \frac{U_s L}{D_T},
\end{equation}
and 
\begin{equation}
    \gamma = \sqrt{D_R\tau_D}.
\end{equation}

To study the first effect of swimming on the macrotransport, we pose a regular expansion for $g$ in the weak-swimming limit characterized by $Pe \ll 1$: 
\begin{equation}
\label{eq:g-Pes-expand}
    g = g^{(0)} + Pe\,  g^{(1)}+Pe^2\,  g^{(2)}+\cdots. 
\end{equation}
Similarly, the orientational moments are written as 
\begin{subequations}
\label{eq:n0-m0-Pes-expand}
    \begin{eqnarray}
        \rho_0 &=& \rho_0^{(0)}+ Pe\, \rho_0^{(1)}+ Pe^2\, n_0^{(2)}+\cdots,\\ 
        \bm{m}_0 &=& \bm{m}_0^{(0)}+ Pe\,  \bm{m}_0^{(1)}+ Pe^2 \,\bm{m}_0^{(2)}+\cdots. 
    \end{eqnarray}
\end{subequations}

At $O(1)$, the swimming motion is absent, and the particles exhibit (passive) Brownian motion. The solution is readily obtained to be 
\begin{equation}
    g^{(0)}(\bx, \bq)  = \frac{1}{|\mathbb{S}|},
\end{equation}
where $S^d$ is the total ``volume'' of the orientation space. As expected, for Brownian particles, the density is uniform, $\rho_0^{(0)} = 1$, and the average drift is zero. 

At $O(Pe)$, one can  show that the density $\rho_0^{(1)} = 0$, and the polar order is governed by 
\begin{subequations}
    \begin{equation}
        - \nabla^2 \bm{m}_0^{(1)} + (d-1)\gamma^2 \, \bm{m}_0^{(1)}=\bm{0}, 
    \end{equation}
    \begin{equation}
        \bn \cdot\nabla\bm{m}_0^{(1)} = \frac{1}{d}\bn, \quad \bx \in \partial \Omega_w.
    \end{equation}
\end{subequations}
The periodic conditions also apply for $\bm{m}_0^{(1)}$. Because the density at $O(Pe)$ vanishes, the $O(Pe)$ average drift is zero.

The first non-zero average drift can be obtained at $O(Pe^2)$; the number density is given by 
\begin{subequations}
    \begin{equation}
        \nabla^2 \rho_0^{(2)} = \nabla\cdot \bm{m}_0^{(1)}, 
    \end{equation}
    \begin{equation}
        \bn\cdot \nabla \rho_0^{(2)} = \bn\cdot\bm{m}_0^{(1)}, \quad \bx \in \partial \Omega_w.
    \end{equation}
\end{subequations}
From \eqref{eq:Ueff-diffusive}, we have the dimensional drift 
\begin{equation}
\label{eq:Ueff-small-Pe}
U^\mathrm{eff} = - Pe^2 \frac{D_T}{|\Omega|}\be_x\cdot \int_{\partial \Omega_t	\cup \partial \Omega_b } \bn \,  \rho_0^{(2)}\, \odiff S + o(Pe^2). 
\end{equation}
To calculate the leading-order drift using the preceding equation, one needs to specify the channel geometry. Nevertheless, a straightforward scaling analysis of \eqref{eq:Ueff-small-Pe} reveals the general behavior that $U^\mathrm{eff}/U_s \sim Pe$ as $Pe \to 0$ for a fixed $\gamma$ (provided that the channel is not fore-aft symmetric). This is an important result of this section: the average drift scaled by the swim speed grows linearly in the weak-swimming limit. Alternatively, scaling the drift by the diffusive speed $D_T/L$, we  have $U^\mathrm{eff}/(D_T/L) \sim Pe^2$. One may interpret the ratio $U^\mathrm{eff}/U_s$ as a rectification efficiency since at most one could align all particles in the same direction via external means and obtains the theoretical maximum of unity:  $U^\mathrm{eff}/U_s = 1$ when $g(\bx, \bq, t) = \rho_0(\bx,t) \delta(\bq-\be_x)$, where $\delta$ is the Dirac delta function. Under channel confinement, particles at the wall tend to point into the wall and their orientations are not strongly aligned with $\be_x$. As a result, the rectification efficiency in periodic channels are often small. 

Next, we consider the displacement field and the resulting dispersion coefficient in the weak swimming limit. The non-dimensional displacement field equation at long times is given by 
\begin{eqnarray}
\label{eq:b-eq-dimensionless}
\nabla\cdot\left(Pe\, \bq\, b -  \nabla b\right) - \gamma^2 \nabla_R^2 b =2  \frac{\partial g}{\partial x} + \left( U^\mathrm{eff} - Pe\, q_x\right)  g, 
\end{eqnarray}
where $U^\mathrm{eff}$ is non-dimensionalized by the diffusive speed $L/D_T$, and $b$ is non-dimensionalized by $L$. Similar to \eqref{eq:g-Pes-expand} and \eqref{eq:n0-m0-Pes-expand}, we may write 
\begin{subequations}
    \begin{eqnarray}
        b &=& b^{(0)} +Pe \,b^{(1)}+Pe^2\, b^{(2)}+\cdots, \\ 
            \rho_1 &=& \rho_1^{(0)}+ Pe \, \rho_1^{(1)}+ Pe^2 \, \rho_1^{(2)}+\cdots,\\ 
        \bm{m}_1 &=& \bm{m}_1^{(0)}+ Pe \, \bm{m}_1^{(1)}+ Pe^2 \, \bm{m}_1^{(2)}+\cdots. 
    \end{eqnarray}
\end{subequations}

At $O(1)$, recall that the particles are passive, and we obtain 
\begin{subequations}
\label{eq:d0-eq-Pes}
    \begin{equation}
       - \nabla^2 b^{(0)}  - \gamma^2 \nabla_R^2 b^{(0)} = 2 \frac{\partial g^{(0)}}{\partial x} =0, 
    \end{equation}
    \begin{equation}
       -\bn\cdot \nabla b^{(0)}=\bn\cdot\be_x\,  g^{(0)}, \quad \bx \in \partial \Omega_w.
    \end{equation}
\end{subequations}
In the absence of swimming, the orientational distribution is uniform, and we have $b^{(0)} =\rho_1^{(0)}/ |\mathbb{S}|$. This means that the polar order vanishes, $\bm{m}_1^{(0)} =\bm{0}$, and we only need to consider the number density. From \eqref{eq:d0-eq-Pes}, we have 
  \begin{subequations}
  \label{eq:rho1-0-weakswim}
    \begin{equation}
       -\nabla^2 \rho_1^{(0)}=0, 
    \end{equation}
    \begin{equation}
       -\bn\cdot \nabla \rho_1^{(0)}=\bn\cdot\be_x, \quad \bx \in \partial \Omega_w.
    \end{equation}
\end{subequations}  
The preceding equations govern the fluctuations in the density field of passive particles, which affects the dispersion coefficient. Equation \eqref{eq:rho1-0-weakswim} is the same as \eqref{eq:rho-1-passive}, which were obtained by setting the swim speed to zero. The density field $\rho_1^{(0)}$ contributes to the dispersion coefficient, see \eqref{eq:D-eff-passive}.  This contribution is a purely passive result. 

It follows from \eqref{eq:d-norm} that the net density must vanish: 
\begin{equation}
  \int_\Omega \rho_1^{(0)} \odiff \bx =0.   
\end{equation}

At $O(Pe)$, the $b$ field is governed by 
\begin{subequations}
    \begin{equation}
        \nabla\cdot\left( \bq\,  b^{(0)} -\nabla b^{(1)} \right) - \gamma^2 \nabla_R^2 b^{(1)}= 2 \frac{\partial g^{(1)}}{\partial x}-q_x g^{(0)},
    \end{equation}
    \begin{equation}
        \bn\cdot\left( \bq \, b^{(0)} - \nabla b^{(1)} \right)= \bn\cdot\be_x g^{(1)},  \quad \bx \in \partial \Omega_w.
    \end{equation}
\end{subequations}
Recalling that $\rho_0^{(1)}=0$ and $b^{(0)}$ does not depend on $\bq$, we conclude that the density at $O(Pe)$, $\rho_1^{(1)}  = \int b^{(1)}\odiff \bq = 0$. The polar order, however, does not vanish; it is governed by 
\begin{subequations}
    \begin{equation}
        \nabla\cdot\left( \frac{\bI}{d} \rho_1^{(0)} - \nabla \bm{m}_1^{(1)}\right) +(d-1)\gamma^2 \bm{m}_1^{(1)} = 2 \frac{\partial \bm{m}_0^{(1)}}{\partial x} - \frac{\be_x}{d},
    \end{equation}
    \begin{equation}
         \bn\cdot\left(\frac{\bI}{d} \rho_1^{(0)} - \nabla \bm{m}_1^{(1)} \right)= \bn\cdot\be_x \bm{m}_0^{(1)},  \quad \bx \in \partial \Omega_w.
    \end{equation}
\end{subequations}
Because the density at $O(Pe)$ vanishes, there is no active contribution to the dispersion at this order (see the last term in \eqref{eq:D-eff-general-simp}). 

To seek the first active contribution to the dispersion coefficient, we have to consider the density field at $O(Pe^2)$. Following the expansion procedure, one can show that $\rho_1^{(2)}$ satisfies
\begin{subequations}
    \begin{equation}
        \nabla\cdot\left(  \bm{m}_1^{(1)} - \nabla \rho_1^{(2)} \right) = 2 \frac{\partial \rho_0^{(2)}}{\partial x} + U^{(2)} - \be_x\cdot\bm{m}_0^{(1)}, 
    \end{equation}
    \begin{equation}
        \bn\cdot\left(  \bm{m}_1^{(1)} - \nabla \rho_1^{(2)} \right) = \bn\cdot\be_x \rho_0^{(2)},  \quad \bx \in \partial \Omega_w.
    \end{equation}
\end{subequations}

Taken together, we obtain the dispersion coefficient as 
\begin{eqnarray}
\label{eq:Deff-weakswim}
     \frac{D^\mathrm{eff}}{D_T} &= &\frac{D_\mathrm{passive}^\mathrm{eff}}{D_T} + \frac{D_0^\mathrm{swim}}{D_T}\left( 1  - d\Bavg{\frac{\partial m_{0,x}^{(1)}}{\partial x} }\right) \nonumber \\ 
     &&+ Pe^2 \Bavg{\frac{\partial \rho_1^{(2)}}{\partial x}}  + o(Pe^2), 
\end{eqnarray}
where we remark that $x$, $\rho_1^{(2)}$, and $\bm{m}_0^{(1)}$ are non-dimensional. Equation \eqref{eq:Deff-weakswim} is the second key result of this section. In the absence of activity ($Pe \equiv 0$), the longitudinal dispersion coefficient is given by the purely passive result of \eqref{eq:D-eff-passive}. Noting that $D^\mathrm{swim}_0/D_T = Pe^2/[\gamma^2d(d-1)]$, we conclude that the first active contribution to $D^\mathrm{eff}/D_T$ is $O(Pe^2)$. Notice that in  asymmetric channels, the undisturbed swim diffusivity $D_0^\mathrm{eff}/D_T$ is corrected by the average $x$ gradient of the polar order $m_{0,1}^{(1)}$. Furthermore, the active density fluctuation also contributes to the dispersion, as given by the first term on the second line of \eqref{eq:Deff-weakswim}. We note that the passive density fluctuation is accounted for in $D_\mathrm{passive}^\mathrm{eff}$.

\subsection{Nearly flat channels}
In this section, we consider nearly flat channels defined by 
\begin{equation}
    y_{\pm} = W \left[ \pm 1+ \epsilon w_\pm (x)\right],
\end{equation}
where $W$ is the half-width of the unperturbed flat channel. The channel modulations  are assumed to be small, i.e., $\epsilon w_{\pm}(x) \ll 1$. Periodicity requires that $w_\pm(x+L) = w_\pm(x)$. In the small-modulation
 limit, we seek an asymptotic solution via 
 \begin{subequations}
 \label{eq:g-d-expand-flat}
 \begin{equation}
       \label{eq:g-ep-expand}
     g = g^{(0)} + \epsilon\,  g^{(1)} + \cdots, 
 \end{equation}
\begin{equation}
    b = b^{(0)} + \epsilon \, b^{(1)} + \cdots.   
\end{equation}
 \end{subequations}
 The unit normal vector at $y = y_{+}$ can be expanded as 
\begin{equation}
\label{eq:n_top_expand}
    \bn\rvert_{y_+} = \be_y + \epsilon\,  \bn^{(1)}+O(\epsilon^2),
\end{equation}
where $\bn^{(1)} = -W w^\prime_{+}(x) \be_x$. Similarly, the unit outward normal to the bottom wall $n\rvert_{y_-} = -\be_y + \epsilon\,  W \, w_-^\prime \be_x+O(\epsilon^2)$.

To study the first effect of small-amplitude modulations, we recast the problem onto the unperturbed domain. From \eqref{eq:Ueff-diffusive}, one can show that the leading-order contribution to the average drift is $O(\epsilon^2)$, $U^\mathrm{eff} = \epsilon^2 U^{(2)}+\cdots$, where 
\begin{equation}
    U^{(2)} =  \frac{D_T}{2L}\int_0^L \left( w_+^\prime \; \rho_0^{(1)}\big\rvert_W -w_-^\prime\;  \rho_0^{(1)}\big\rvert_{-W} \right)\odiff x. 
\end{equation}
In the above, $(\cdot)\rvert_{y^*}$ is a shorthand for evaluating the expression at $y=y^*$. Similarly, the effective dispersion coefficient admits the expansion $D^\mathrm{eff} = D_0^\mathrm{eff} +\epsilon^2 D^{(2)}+\cdots$, where 
\begin{eqnarray}
     D^{(2)} =&& \frac{U_s\delta^2 W}{d-1} \int_0^L \left( w_+^\prime \; m_{0,x}^{(1)}\big\rvert_W -w_-^\prime \; m_{0,x}^{(1)}\big\rvert_{-W}  \right) \odiff x  \nonumber\\ 
     && - D_T W \int_0^L \left( w_+^\prime \; \rho_{1}^{(1)}\big\rvert_W -w_-^\prime \; \rho_{1}^{(1)}\big\rvert_{-W}  \right) \odiff x.  
\end{eqnarray}
It is clear that for a flat channel, $\epsilon \equiv 0$, the average drift vanishes and the effective dispersion coefficient is $D_0^\mathrm{eff} = D_T+D^\mathrm{swim}_0$.

To calculate $U^{(2)}$ and $D^{(2)}$, we need the fields $g^{(1)}$ and $b^{(1)}$ defined in \eqref{eq:g-d-expand-flat}. We first expand $g$ at $y=y_+$ about the unperturbed boundary $y=W$ in a Taylor series, which gives 
\begin{equation}
\label{eq:domain-perturb}
    g\rvert_{y_+} = g\rvert_W + \frac{\partial g}{\partial y}\Big\rvert_W \epsilon \,w_+W  + \cdots,
\end{equation}
 Combining \eqref{eq:g-ep-expand} and  \eqref{eq:domain-perturb}, we have 
\begin{eqnarray}
\label{eq:g_top_expand}
    g\rvert_{y_+} = g^{(0)}\big\rvert_W  +\epsilon\left( g^{(1)}\big\rvert_W + \frac{\partial g^{(0)}}{\partial y}\Big\rvert_W\,  w_+\, W\right) +O(\epsilon^2). 
\end{eqnarray}
 For the bottom wall, $y=y_{-}$, a similar expansion can be written. At $O(\epsilon^p)$, $p=0,1,2,\cdots$, Eq. \eqref{eq:g-eq} remains unchanged, which gives 
\begin{equation}
\label{eq:g0-flat}
     \nabla\cdot\left(U_s\bq g^{(p)} - D_T \nabla g^{(p)}\right) - D_R \nabla_R^2 g^{(p)}=0. 
\end{equation}
At $O(1)$, the channel is flat, and the no-flux condition becomes 
\begin{equation}
\label{eq:bc-g0-glat}
     U_s q_y g^{(0)} - D_T \frac{\partial g^{(0)}}{\partial y}=0, \quad y = \pm W. 
\end{equation}
Notice that the boundary condition is now imposed at the unperturbed flat walls, which is the purpose of the domain perturbation analysis. Equations \eqref{eq:g0-flat} and \eqref{eq:bc-g0-glat} specify the dynamics of ABPs in a flat channel; $g^{(0)}$ is invariant in $x$. 

At $O(\epsilon)$, the no-flux condition is 
\begin{eqnarray}
\label{eq:g-flat-1-bc}
     &&U_s q_y g^{(1)}- D_T \left( \frac{\partial g^{(1)}}{\partial y} + 
 w_\pm \; W \frac{\partial^2 g^{(0)}}{\partial y^2}\right) \nonumber \\ 
 && = w^\prime_\pm\; W \left( U_s q_xg^{(0)} - D_T \frac{\partial g^{(0)}}{\partial x} \right),\quad y= \pm W, 
\end{eqnarray}
where the last term vanishes because $\partial g^{(0)}/\partial x=0$. Following the above procedure, one may obtain the governing equations for $b^{(0)}$ and $b^{(1)}$, from which the density and polar order of the displacement field can be calculated. 

% Taking the first two orientational moments of \eqref{eq:g-flat-1-bc}, we obtain 
% \begin{eqnarray}
%     &&U_s m_{0,y}^{(1)} - D_T \frac{\partial n_0^{(1)}}{\partial y} = D_T w_\pm \; W \frac{\partial^2 n_0^{(0)}}{\partial y^2},\\ 
%     && U_s Q_{0,yx}^{(1)}-D_T \frac{\partial m_{0,x}^{(1)}}{\partial y} = U_sw_\pm^\prime W  \tilde{Q}_{0,xx}^{(0)}, 
% \end{eqnarray}
% where we have used the fact that $m_{0,x}^{(0)}=0$ \highred{(How to prove this?, what about $Q_{xx}$, maybe move the moments to the example section where it is actually used.)}. 

\begin{figure}[htb]
\includegraphics[width=2.9in]{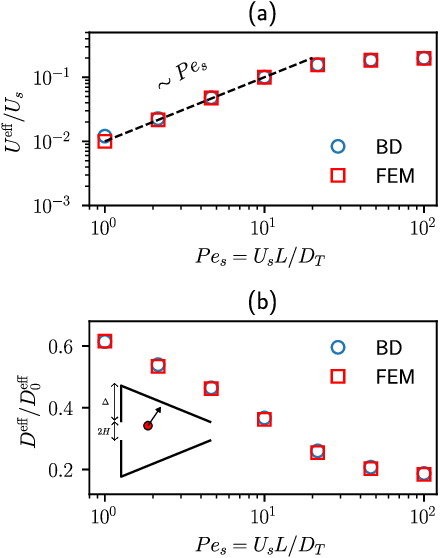}
\caption{\label{fig:Ueff-Deff} (a) The dimensionless average drift $U^\mathrm{eff}/U_s$ as a function of the swimming P\'eclet number ($Pe_s=U_sL/D_T)$.  (b) The dimensionless effective dispersion coefficient $D^\mathrm{eff}/D_0^\mathrm{eff}$ as a function of the swimming P\'eclet number ($Pe_s=U_sL/D_T)$. For both (a) and (b), $\gamma=1$ and the channel unit cell is shown in the inset of (b). The geometry of the channel is characterized by $H/L=0.1$ and $\Delta/L=0.4$, where $L$ is the (horizontal) length of the unit cell. No external forces or fields are considered. Square symbols denote theoretical results obtained from solving the macrotransport equations using a standard finite element method (FEM). Circles are results from BD simulations. The dashed line in (a) shows a linear dependence of $U^\mathrm{eff}/U_s$ on $Pe_s$ (see Sec. \ref{subsec:weak-swim}). }
\end{figure}

\section{Numerical examples}
\label{sec:num-examples}

To provide a direct numerical validation of our macrotrans-
port equations, we compare the macrotransport coefficients
obtained from the macrotransport theory with those obtained
from BD simulations. We describe below the rectified transport and long-time dispersion dynamics of ABPs in asymmetric confinement. The mean and displacement equations are solved using a standard finite element method (FEM) implemented in \texttt{FreeFem++}~\cite{hecht2012new}. In BD simulations, the corresponding Langevin equations of motion of an ABP is integrated in time using a standard Euler-Maruyama scheme. The drift coefficient is obtained by fitting the slope of the mean-displacement curve at long times. The implementation details of FEM and BD are presented in the supplementary material.

\begin{figure}[htb]
\includegraphics[width=3in]{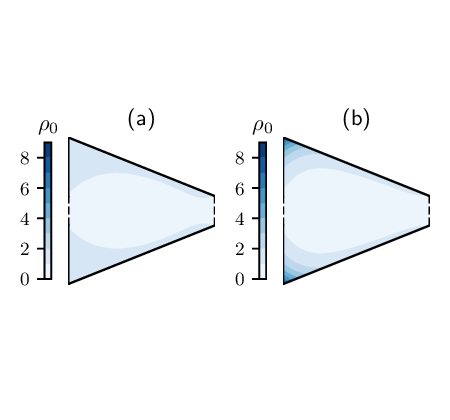}
\caption{\label{fig:density} The number density distribution ($\rho_0$) in the unit cell for (a) $Pe=1$ and (b) $Pe=10$ obtained from FEM. For both panels, we have $\gamma=1$. The geometric parameters of the channel are given in Fig.~\ref{fig:Ueff-Deff}. The channel openings on the left and right are shown by dashed lines, while the no-flux walls are shown by solid lines. }
\end{figure}

\subsection{An asymmetric channel}

For an ABP in an asymmetric channel, the non-dimensional average and displacement equations are, respectively, \eqref{eq:g-non-dim} and \eqref{eq:b-eq-dimensionless}. Once the macrotransport equations are solved, we calculate the average drift and effective dispersion coefficient using \eqref{eq:Ueff-eq-general} and \eqref{eq:Deff-eq-general}. In Fig.~\ref{fig:Ueff-Deff}(a), we plot the average drift scaled by the swim speed as a function of the swimming P\'eclet number ($Pe$). The dimensionless drift increases with increasing $Pe$ (or swim speed) before it saturates at large $Pe$. As derived in Sec.~\ref{subsec:weak-swim} using an asymptotic analysis, we have $U^\mathrm{eff}/U_s \sim Pe$ for slow swim speeds. Such a linear scaling has been inferred from fitting BD simulation data~\cite{Ghosh2013}. In Fig.~\ref{fig:Ueff-Deff}(b), we plot the longitudinal dispersion coefficient scaled by $D_0^\mathrm{eff}$ (recall that $D_0^\mathrm{eff} =D_T+D_0^\mathrm{swim}$) as a function of $Pe$. As the swim speed increases, the reduction in dispersion is more pronounced because the effective run length is reduced as particles become more confined ($\ell/W =U_s\tau_R/W \gg 1$, where $W$ is the characteristic width of the channel). We note that the confinement-induced hindrance in effective diffusion has been observed~\cite{ao2014active,modica2023boundary}. For both the drift and dispersion coefficient, good agreement between the macrotransport theory (squares) and BD (circles) simulations are observed.

In Fig.~\ref{fig:density}, we show the number density distribution of the average field for (a) $Pe=1$ and (b) $Pe=10$. The number density field is obtained by taking the zeroth moment of the FEM solution of $g$.  That is, we calculate the integral ($\rho_0 = \int_{\mathbb{S}} g \odiff\bq$) using the trapezoidal rule. For low activity (e.g. $Pe=1$), we see that the density gradient is very weak. When activity is high (e.g. $Pe=10$), the boundary accumulation of ABPs can be seen clearly from the steep density gradients near the top and bottom corners. As shown by our theory, the net diffusive flux gives the average drift. From Fig.~\ref{fig:density}(b), we see that the density is higher on the left side compared with that on the right. Therefore, the net motion is towards the right, i.e. $U^\mathrm{eff}>0$.

\subsection{An asymmetric substrate potential}
Consider an ABP subject to an asymmetric substrate potential given by $V(x) = V_0[\sin(2\pi x/L) + \sin(4\pi x/L)/4]$ in two dimensions.\cite{reichhardt2013active} The force exerted on the particle due to the potential is $\bF = - \partial V/(\partial x)\be_x$. Note that the potential has a periodicity of $L$ and the force biases the particles towards the negative $x$ direction. Besides $Pe$ and $\gamma$ introduced in Sec.~\ref{subsec:weak-swim}, the system is also governed by a P\'eclet number of the external potential, which is defined as $Pe_V= V_0\pi/(\zeta D_T)$. 

\begin{figure}
\includegraphics[width=2.9in]{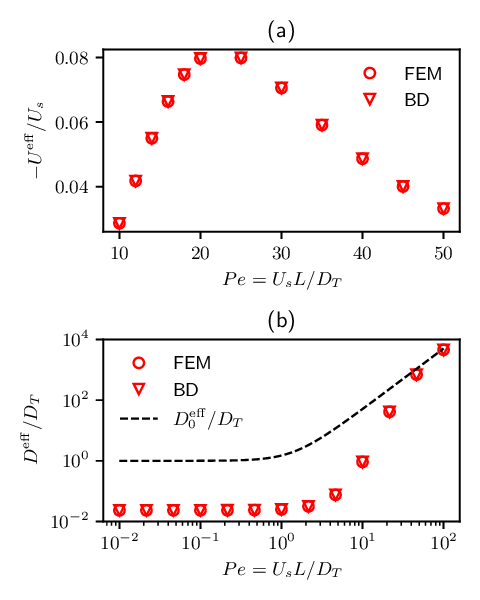}
\caption{\label{fig:Ueff-Deff-potential}Macrotransport coefficients of ABPs in a one-dimensional periodic substrate potential. (a) The dimensionless average drift $-U^\mathrm{eff}/U_s$ as a function of the swimming P\'eclet number ($Pe_s=U_sL/D_T)$. We note that $U^\mathrm{eff}$ is negative. (b) The dimensionless effective dispersion coefficient $D^\mathrm{eff}/D_T$ as a function of the swimming P\'eclet number ($Pe_s=U_sL/D_T)$. For both (a) and (b), $\gamma=1$ and $Pe_V=10$. Circle symbols denote theoretical results obtained from solving the macrotransport equations using FEM. Triangles are results from BD simulations. The free-space result is shown by the dashed line.}
\end{figure}

We show the macrotransport coefficients along the $x$ direction in Fig.~\ref{fig:Ueff-Deff-potential}. The dimensionless average drift as a function of $Pe$ is plotted in Fig.~\ref{fig:Ueff-Deff-potential}(a). Because the drift is negative, the quantity $-U^\mathrm{eff}/U_s$ is plotted. In Fig.~\ref{fig:Ueff-Deff-potential}(b), we plot the dimensionless dispersion coefficient ($D^\mathrm{eff}/D_T$) as a function of $Pe$. Good agreement between the theory (FEM) and BD simulations are observed. Different from a hard wall that can exert a force as large as necessary to prevent a particle from moving beyond the no-flux boundary, a substrate potential is soft. When the swim force ($\zeta U_s$) exceeds the potential force, the effect of the potential becomes weak. For a fixed $Pe_V$, we therefore observe that the net ratchet flux vanishes and the effective diffusivity approaches $D_0^\mathrm{eff}$ as $Pe\to \infty$. As shown in Fig.~\ref{fig:Ueff-Deff-potential}(a), the dimensionless average drift varies non-monotonically as a function of $Pe$. The maximum rectification efficiency (defined as $-U^\mathrm{eff}/U_s)$ is achieved when the swim force is comparable with the force due to the potential. The force due to the substrate potential suppresses the effective dispersion. As a result, we see in Fig.~\ref{fig:Ueff-Deff-potential}(b) that $D^\mathrm{eff}$ is lower than $D_0^\mathrm{eff}$ for finite $Pe$. In the large-$Pe$ limit, $D^\mathrm{eff}$ approaches that of the free-space value since the substrate force becomes sub-dominant.

\section{Discussion}
\label{sec:discussion}
In this paper, we have developed a macrotransport theory for active Brownian particles in spatially periodic domains. By decomposing the position vector into the cell vector and a local position vector, we derived a pair of macrotransport equations that characterize the long-time transport dynamics of active particles.  Our macrotransport theory can be treated as a generalized Taylor dispersion theory (GTDT) for periodic geometries. In the language of GTDT, the cell index is the global space, while the local position vector and the orientation vector constitute the local space. Leveraging this generic framework, we derived macrotransport equations for active particles in corrugated channels, spatially periodic external fields, and activity landscapes. Under geometric confinement, while the focus was on corrugated periodic channels, the theoretical model can readily accommodate the macrotransport processes of active particles in periodic porous media~\cite{Saintillan2019,dehkharghani2019bacterial,Kumar2022,modica2023boundary}.

For active particles confined in a corrugated periodic channel in the absence of external forces and fields, it is known that the active Brownian motion under a no-flux boundary condition can be rectified provided that the unit cell of the channel is fore-aft asymmetric. Complementing previous works on BD simulations and experiments, our approach provides a statistical mechanical description of the rectification mechanism. Regardless of activity, the average drift or rectified flux is given by the net diffusive flux along the channel. For passive Brownian motion, the density is homogeneous, and no net flux is generated. In the case of active particles, their activity coupled with confinement can support a non-zero density gradient. Ultimately, such a density gradient leads to rectified transport. This autonomous rectification reflects the non-equilibrium nature of active systems.  In addition to the average drift, the macrotransport theory allows us to characterize the longitudinal dispersion coefficient of active particles. While different boundary conditions may be incorporated in the GTDT, in this work we focused on the case of a no-flux boundary condition. For the cases of periodic potentials or fields in the absence of channel confinement, only periodic conditions due to the geometry are needed.

In modeling the effective longitudinal dynamics of (passive) Brownian motion under spatially varying confinement such as a corrugated channel, the Fick-Jacobs (FJ) theory has been widely adopted~\cite{jacobs1935diffusion,zwanzig1992diffusion,reguera2001kinetic,Kalinay05,reguera2006entropic,kalinay2008approximations,burada2007biased}. The FJ theory reduces the problem of diffusion in two- or three-dimensional channels to an effective one-dimensional (1D) problem along the channel axis. This is done by averaging over the cross-section of the channel, which often relies on the assumption that the cross-section varies slowly.  In doing so, the spatial variation of the channel is incorporated into an effective diffusion coefficient that depends on the channel geometry.  Often, the dependence of the diffusion coefficient in the 1D equation on the channel geometry is defined heuristically~\cite{reguera2001kinetic,reguera2006entropic}. Recently, the FJ theory has been extended to accommodate active Brownian motion~\cite{modica2023boundary} by treating active particles as (passive) Brownian particles at a higher effective temperature. Not surprisingly, the FJ theory deviates from the BD simulation results when active particles are sufficiently confined. Despite the discrepancy, the simplicity of FJ theory allows one to obtain a quick estimate of the dispersion behavior. In contrast, we emphasize that our macrotransport theory does not restrict the geometry of the channel unit cell. Said differently, the macrotransport theory works equally well regardless of the degree of confinement that the active particles experience. As such, it does not impose a restriction on the variation of the channel walls or on the run length (or activity) of the active particles. Because the macrotransport equations depend on the full position-orientation space coordinates (unlike the 1D equation in FJ theory), their application often requires numerical solutions of partial differential equations, as demonstrated in Fig.~\ref{fig:Ueff-Deff}.

To develop the macrotransport theory, several simplifying assumptions have been made. The governing equations are written for a single point-sized particle in the absence of particle-particle and particle-wall hydrodynamic interactions. Even for single particle dynamics, finite-size effects may become important if the width of the channel is comparable to the size of the particle. In this case, one may still employ the single-particle Smoluchowski equation, but the appropriate hydrodynamic resistance tensors need to be considered. For active particle systems with finite densities, interactions among particles---whether hydrodynamic or otherwise---may play a role in dictating the macrotransport dynamics. In principle, one can make use of the multi-particle Smoluchowski equation to develop a statistical mechanical description~\cite{peng2022trapped}. However, the high-dimensionality of such equations poses a computational challenge. As such, particle-based simulation methods may be more convenient for investigating the transport processes of semi-dilute and dense systems. In the case of passive-active mixtures, the macrotransport dynamics under asymmetric confinement becomes more intricate.\cite{Ghosh2013,lu2017ratchet,debnath2020enhanced,wang2020different,wang2023controlling} Recent work suggests that, depending on the channel and particle geometry, the drift of the mixture can be in either direction of the channel.\cite{wang2023controlling} 

In this work, the asymmetric confining boundaries (e.g. corrugated channel) are not allowed to move. To obey momentum balance, active particles subject to a no-flux boundary condition exhibit a rectified flux. In the case of an asymmetric passive object immersed in a bath of active particles, the object itself can exhibit directed linear motion.\cite{angelani2010geometrically,RR,belan2021active,dolai2024shape} Similarly, a gear with asymmetric teeth immersed in an active bath can exhibit directed rotational motion.\cite{Angelani,Leonardo,Sokolov,xu2021rotation,jerez2020dynamics} Our work provides a mathematical foundation to develop statistical mechanical theories to characterize the transport mechanics of immersed objects in an active bath. Theoretical models can be useful in identifying limits in the extent of useful work that can be extracted from the active Brownian motion of the bath. In addition,  it will be useful for the development of design principles for activity-induced transport.\cite{pietzonka2019autonomous,peng_zhou_brady_2022} 

% \begin{acknowledgments}
% We wish to acknowledge the support of the author community in using
% REV\TeX{}, offering suggestions and encouragement, testing new versions,
% \dots.
% \end{acknowledgments}
\section*{Supplementary Material}
In the supplementary material, we provide details of mathematical derivations, FEM and BD simulations. 

\section*{Data Availability Statement}
The data that support the findings of this study are available within the article.

\bibliography{refs}% Produces the bibliography via BibTeX.

\end{document}